\RequirePackage{fix-cm}
\documentclass[twocolumn,epjc3]{svjour3}  
\smartqed  
\RequirePackage{graphicx}
\RequirePackage{comment}
\usepackage{color}
\usepackage{lineno}
\usepackage{url}

\newcommand{\DBD}{0$\nu$DBD}
\newcommand{\mbb}{m$_{\beta\beta}$}

\newcommand{\TEHT}{$^{130}\mathrm{Te}$}
\newcommand{\TEO}{$\mathrm{TeO}_2$}
\newcommand{\CuoreZ}{CUORE-0}
\newcommand{\cuoreZlimit}{T$_{1/2}^{0\nu}>$4.0$\times$10$^{24}$\,y at 90$\%$ C.L.}
\newcommand{\cuoreZlimitMass}{m$_{\beta\beta}<$270-760\,meV}

\newcommand{\cuorelimit}{T$_{1/2}^{0\nu}$($^{130}$Te)$>$9.5$\times$10$^{25}$\,y at 90$\%$ C.L.}
\newcommand{\cuorelimitMass}{m$_{\beta\beta}<$50-130\,meV}

\newcommand{\Se}{$^{82}\mathrm{Se}$}
\newcommand{\qvalue}{2997.9$\pm$0.3\,keV}
\newcommand{\duenu}{T$_{1/2}^{2\nu}$=(9.6$\pm$0.1(stat)$\pm$1.0(syst))$\times$10$^{19}$\,y}

\newcommand{\TL}{$^{208}\mathrm{Tl}$}

\newcommand{\CupidZ}{CUPID-0}
\newcommand{\EnrZnSe}{Zn$^{82}$Se}

\newcommand{\MO}{M\Ohm}

\newcommand{\Ohm}{\un\Omega}

\providecommand*{\un}[1]{\ensuremath{\mathrm{~#1}}}
\journalname{Eur. Phys. J. C}

\begin{document}

\title{First array of enriched Zn$^{82}$Se bolometers to search for double beta decay}
\author{D.~R.~Artusa\thanksref{LNGS,USC}
\and A.~Balzoni\thanksref{Roma,INFNRoma}
\and J.~W.~Beeman\thanksref{LBNL}
\and F.~Bellini\thanksref{Roma,INFNRoma}
\and M.~Biassoni\thanksref{INFNMiB}
\and C.~Brofferio\thanksref{MIB,INFNMiB}
\and A.~Camacho\thanksref{Legnaro}
\and S.~Capelli\thanksref{MIB,INFNMiB}
\and L.~Cardani\thanksref{e1,INFNRoma,Princeton}
\and P.~Carniti\thanksref{MIB,INFNMiB}
\and N.~Casali\thanksref{Roma,INFNRoma}
\and L.~Cassina\thanksref{MIB,INFNMiB}
\and M.~Clemenza\thanksref{MIB,INFNMiB}
\and O.~Cremonesi\thanksref{INFNMiB}
\and A.~Cruciani\thanksref{Roma,INFNRoma}
\and A.~D'Addabbo\thanksref{LNGS}
\and I.~Dafinei\thanksref{INFNRoma}
\and S.~Di~Domizio\thanksref{Genova,INFNGenova}
\and M.~L.~di~Vacri\thanksref{LNGS}
\and F.~Ferroni\thanksref{Roma,INFNRoma}
\and L.~Gironi\thanksref{MIB,INFNMiB}
\and A.~Giuliani\thanksref{CNRS,DiSAT}
\and C.~Gotti\thanksref{MIB,INFNMiB}
\and G.~Keppel\thanksref{Legnaro}
\and M.~Maino\thanksref{MIB,INFNMiB}
\and M.~Mancuso\thanksref{CNRS,DiSAT,e2}
\and M.~Martinez\thanksref{Roma,INFNRoma}
\and S.~Morganti\thanksref{INFNRoma}
\and S.~Nagorny\thanksref{GSSI}
\and M.~Nastasi\thanksref{MIB,INFNMiB}
\and S.~Nisi\thanksref{LNGS}
\and C.~Nones\thanksref{CEA}
\and F.~Orio\thanksref{INFNRoma}
\and D.~Orlandi\thanksref{LNGS}
\and L.~Pagnanini\thanksref{GSSI}
\and M.~Pallavicini\thanksref{Genova,INFNGenova}
\and V.~Palmieri\thanksref{Legnaro}
\and L.~Pattavina\thanksref{LNGS}
\and M.~Pavan\thanksref{MIB,INFNMiB}
\and G.~Pessina\thanksref{INFNMiB}
\and V.~Pettinacci\thanksref{Roma,INFNRoma}
\and S.~Pirro\thanksref{LNGS}
\and S.~Pozzi\thanksref{MIB,INFNMiB}
\and E.~Previtali\thanksref{INFNMiB}
\and A.~Puiu\thanksref{MIB,INFNMiB}
\and C.~Rusconi\thanksref{INFNMiB,DiSAT}
\and K.~Sch\"affner\thanksref{GSSI}
\and C.~Tomei\thanksref{INFNRoma}
\and M.~Vignati\thanksref{INFNRoma}
\and A.~Zolotarova\thanksref{CEA}
}

\institute{INFN - Laboratori Nazionali del Gran Sasso, Assergi (L'Aquila) I-67010 - Italy\label{LNGS}
\and
Department of Physics  and Astronomy, University of South Carolina, Columbia, SC 29208 - USA\label{USC}
\and
Dipartimento di Fisica, Sapienza Universit\`{a} di Roma, Roma I-00185 - Italy \label{Roma}
\and
INFN - Sezione di Roma, Roma I-00185 - Italy\label{INFNRoma}
\and
Materials Science Division, Lawrence Berkeley National Laboratory, Berkeley, CA 94720 - USA\label{LBNL}
\and
Dipartimento di Fisica, Universit\`{a} di Milano-Bicocca, Milano I-20126 - Italy\label{MIB}
\and
INFN - Sezione di Milano Bicocca, Milano I-20126 - Italy\label{INFNMiB}
\and
INFN - Laboratori Nazionali di Legnaro, Legnaro (Padova) I-35020 - Italy \label{Legnaro}
\and
Physics Department - Princeton University, Washington Road, 08544, Princeton - NJ, USA\label{Princeton}
\and
Dipartimento di Fisica, Universit\`{a} di Genova, Genova I-16146 - Italy\label{Genova}
\and
INFN - Sezione di Genova, Genova I-16146 - Italy\label{INFNGenova}
\and
CSNSM, Univ. Paris-Sud, CNRS/IN2P3, Universit\'e Paris-Saclay, 91405 Orsay, France\label{CNRS}
\and
DiSAT, Universit\`{a} dell'Insubria, 22100 Como, Italy\label{DiSAT}
\and
CEA-Saclay, DSM/IRFU, 91191 Gif-sur-Yvette Cedex, France\label{CEA}
\and
INFN - Gran Sasso Science Institute, 67100, L'Aquila - Italy\label{GSSI}
}

\thankstext{e1}{e-mail: laura.cardani@roma1.infn.it}
\thankstext{e2}{Present address: Max-Planck-Institut f\"ur Physik, 80805, M\"unchen, Germany}


\date{Received: date / Accepted: date}

\maketitle

\begin{abstract}
The R$\&$D activity performed during the last years proved the potential of ZnSe scintillating bolometers to the search for neutrino-less double beta decay,
motivating the realization of the first large-mass experiment based on this technology: \CupidZ.
The isotopic enrichment in $^{82}$Se, the \EnrZnSe\ crystals growth, as well as the light detectors production have been accomplished, and the experiment is now in construction at Laboratori Nazionali del Gran Sasso (Italy).
In this paper we present the results obtained testing the first three \EnrZnSe\ crystals operated as scintillating  bolometers, and we prove that their performance in terms of energy resolution, background rejection capability and intrinsic radio-purity complies with the requirements of \CupidZ.

\end{abstract}

\section{Introduction}
\label{intro}
Neutrino-less double beta decay (\DBD) is a hypothesized nuclear transition in which a nucleus decays emitting only two electrons.
This process cannot be accommodated in the Standard Model, as the absence of emitted neutrinos would violate the lepton number conservation. 
For this reason, its prized observation would have several implications for particle physics, astrophysics and cosmology.
According to the majority of theoretical frameworks, for \DBD\ to happen neutrinos must be Majorana particles. This means that, in contrast to all the other known fermions, they must coincide with their own antiparticles.
In this scenario, the half-life of the process T$_{1/2}^{0\nu}$ is determined by the effective Majorana neutrino mass \mbb, a linear superposition of the three neutrinos mass eigenstates weighted by the elements of the first row of the PMNS neutrino mixing matrix and including two Majorana phases, providing the dependance: T$_{1/2}^{0\nu} \propto 1/m_{\beta\beta}^2$.

Current lower limits on T$_{1/2}^{0\nu}$ exceed 10$^{25}$\,y for several possible emitters, corresponding to upper bounds on \mbb\ of a few hundreds of meV~\cite{Dell'Oro:2016dbc}.
Moreover, recent measurements made by the KamLAND-Zen collaboration using a xenon-loaded liquid scintillator allowed to set a limit on T$_{1/2}^{0\nu}$($^{136}$Xe) of 1.1$\times$10$^{26}$\,y at 90$\%$ C.L., corresponding to \mbb\ $<$ 60-161\,meV~\cite{KamLAND-Zen:2016pfg}.

Among the several experimental approaches proposed for the search of \DBD, cryogenic calorimeters~\cite{Fiorini:1983yj} (historically also called bolometers) stand out for the possibility of achieving excellent energy resolution (of the order of 0.1$\%$), efficiency ($>$80$\%$) and intrinsic radio-purity.
Moreover, the crystals that are operated as bolometers can be grown starting from most of the \DBD\ emitters, enabling the test of different nuclei.

The \CuoreZ\ experiment proved the potential of the bolometric technique by reaching an unprecedented sensitivity on the half-life of \TEHT~\cite{Alfonso:2015wka}.
The \TEHT\ source was embedded in 52 \TEO\ bolometers (for a total mass of 39\,kg), operated between March 2013 and March 2015 at Laboratori Nazionali del Gran Sasso (LNGS, Italy).
The combination of the 9.8\,kg($^{130}$Te)$\cdot$y exposure collected by \CuoreZ\ with the 19.75\,kg($^{130}$Te)$\cdot$y of its ancestor, Cuoricino, allowed to set a new limit on the T$_{1/2}^{0\nu}$($^{130}$Te):  \cuoreZlimit, corresponding to \cuoreZlimitMass. 

The evolution of \CuoreZ, named CUORE~\cite{Artusa:2014lgv}, will start operation before the end of 2016 with 988 \TEO\ bolometers, pointing to a sensitivity of \cuorelimit\ (\cuorelimitMass). 
This remarkable sensitivity will allow to touch, but not to fully explore, the m$_{\beta\beta}$ region of 10-50\,meV, corresponding to the inverted mass hierarchy of neutrinos (see Refs~\cite{Dell'Oro:2016dbc,Vissani3,Vissani4,Vissani10} and references therein for a complete discussion on the relationship between the \DBD\ and the neutrino mass hierarchy).
The entire coverage of this region, indeed, requires a level of sensitivity of about 10$^{27}$-10$^{28}$\,y, depending on the isotope~\cite{Artusa:2014wnl}.
The CUPID (CUORE Upgrade with Particle IDentification~\cite{Wang:2015taa,Wang:2015raa}) interest group is defining the technological upgrades that will allow to reach this target with a CUORE-size bolometric detector.
Such an ambitious goal poses a stringent requirement on the background, that must be close to zero at the tonne$\cdot$year level of exposure.

The experience gained during the years of R$\&$D activity for CUORE allowed to determine that the ultimate limit in the background suppression resides in the presence of $\alpha$-decaying isotopes located in the detector structure.
Thus, the main breakthrough of CUPID with respect to CUORE will be the addition of independent devices to measure the light signals emitted either from Cherenkov radiation (in \TEO~\cite{TabarellideFatis:2009zz}) or from scintillation (in scintillating crystals~\cite{Pirro:2005ar} like ZnSe~\cite{Arnaboldi:2010jx,Beeman:2013vda}, ZnMoO$_4$~\cite{Gironi:2010hs,Beeman:2012ci,Beeman:2012jd,Beeman:2011bg,Cardani:2013mja,Armengaud:2015hda,Berge:2014bsa}, Li$_2$MoO$_4$~\cite{Cardani:2013dia,Bekker:2014tfa} and many others). The different light emission of electrons and $\alpha$ particles will enable event-by-event rejection of $\alpha$ interactions, suppressing the overall background in the region of interest for \DBD\ of at least one order of magnitude.
During the last years, several R$\&$D activities demonstrated that the detection of Cherenkov light emitted by electrons crossing the \TEO\ crystals of CUORE can provide an effective $\alpha$ background rejection~\cite{Casali:2014vvt}. Despite the encouraging results~\cite{Casali:2016luq,Casali:2015gya,Cardani:2015tqa,Battistelli:2015vha,Schaffner:2014caa,Gironi:2016nae,Biassoni:2015eij,Willers:2014eoa}, 
none of these activities is yet mature for a next-generation experiment. 
A different approach was followed by the LUCIFER~\cite{Beeman:2013sba} and LUMINEU~\cite{Barabash:2014una} collaborations, that tested different scintillating bolometers and proved that the read-out of the much more intense scintillation light (a few keV or tens of keV depending on the crystal, with respect to $\sim$100\,eV produced by Cherenkov radiation) simplifies the suppression of the $\alpha$ background. 
Moreover, choosing crystals containing high Q-value \DBD\ emitters such as $^{82}$Se and $^{100}$Mo provides a natural suppression of the $\gamma$ background produced by the environmental radioactivity, that drops above the 2615\,keV $\gamma$-line of \TL.
The encouraging results obtained with large ZnSe bolometers motivated the endeavor in realizing a first large mass demonstrator of the scintillating bolometers technology: \CupidZ. 
A second phase of \CupidZ, not described in this paper, will include also enriched Li$_2$$^{100}$MoO$_4$ and Zn$^{100}$MoO$_4$ crystals, which showed excellent performance as scintillating bolometers.
\CupidZ\ aims at proving that the interactions due to $\alpha$ particles can be efficiently suppressed, providing a path to ensure the achievement of the zero-background level required by CUPID.


\section{CUPID-0}
\label{sec:cupid}
The \CupidZ\ detector will consist of 30 \EnrZnSe\ bolometers arranged in 5 towers, for a total mass of 13.2\,kg (7.0\,kg of \Se).
The Q-value of \Se\ (\qvalue~\cite{Lincoln:2012fq}) is among the highest, and the half-life of the 2$\nu\beta\beta$ mode is long enough (\duenu~\cite{2nuSeHalfLife}) to prevent 2$\nu\beta\beta$ background contributions in the region of interest. 
To date, all the crystals (\EnrZnSe\ and Ge light detectors) were grown and delivered to LNGS. The assembly of the \CupidZ\ is expected to be completed before the end of Summer 2016.

The Se powder was enriched at URENCO Stable Isotope Group (Netherlands) to overcome the rather poor natural isotopic abundance of \Se\ (8.73$\%$~\cite{SeIsotopicAbundance}).
We measured the radio-purity of the 96.3$\%$ enriched powder with HP-Ge spectroscopy and, since we found no evidences of contaminations, we set upper limits of 61\,$\mu$Bq/kg, 110\,$\mu$Bq/kg and 74\,$\mu$Bq/kg at 90$\%$ C.L. for $^{232}$Th, $^{238}$U and $^{235}$U, respectively~\cite{Beeman:2015xjv}.
We measured with the same sensitivity also Zn metal, and found no signatures of  $^{232}$Th, $^{238}$U and $^{235}$U.

The \EnrZnSe\ synthesis, as well as the crystals growth were performed at the Institute for Scintillation Materials (ISMA, Kharkov, Ukraine).
To prevent contaminations, the synthesis and the subsequent powder purification treatments were performed in Argon atmosphere using tools/containers selected according to their radio-purity. 
After the test run described in this paper, all the 95.4$\%$ enriched \CupidZ\ crystals were cut, shaped and polished in order to obtain bolometers with 4.4\,cm diameter and 5.5\,cm height, for an average mass of 440.5\,g.
These operations were carried out in a clean room at LNGS following the protocols for radio-purity described in Ref.~\cite{ioanprod}.
The procedures that we developed for the \EnrZnSe\ synthesis and purification allowed to reach a yield of 98$\%$, with 2$\%$ irrecoverable loss of enriched Se. The yield of the crystals production was reduced by losses of material evaporated during the crystal growth. The crystals presented in this paper were grown using graphite crucible, with an irrecoverable loss of material of about 15$\%$. The growth procedure was then optimized in order to reduce the material loss down to 2$\%$ using glassy graphite crucible. The fraction of material discarded during the manufacturing process (about 40$\%$) can be reprocessed to obtain other \EnrZnSe\ crystals. More details about the synthesis, crystal growth and processing can be found in Ref~\cite{ZnSeGrowth}.

\begin{figure}[thb]
\begin{centering}
\includegraphics[width=\columnwidth]{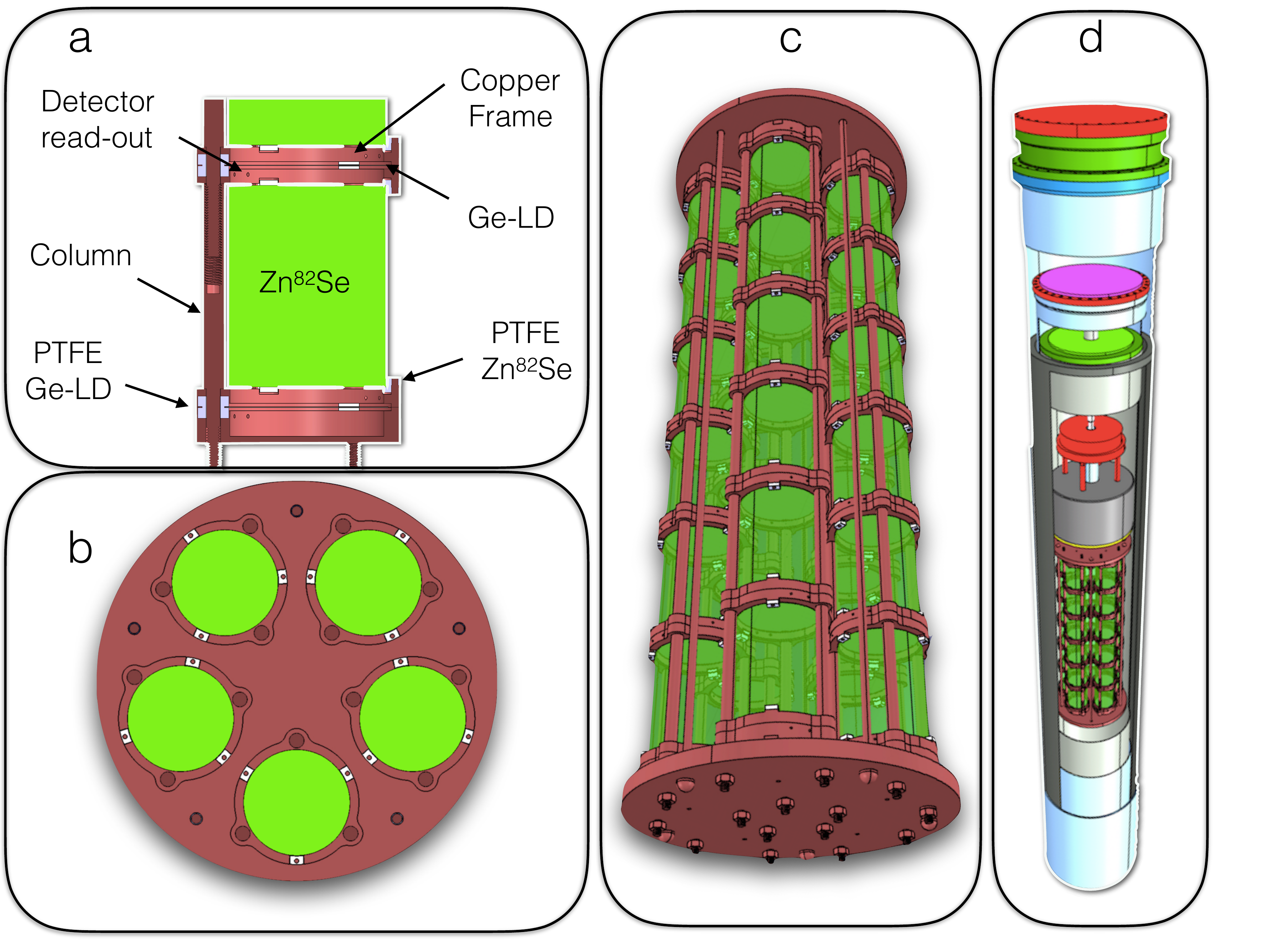}
\caption{a) Lateral view of a single module: the \EnrZnSe\ is placed between two light detectors (Ge-LD), not visible in this scheme because of their small thickness (about 170\,$\mu$m); all the detectors are held in the copper structure using PTFE elements. Top (b) and 3D (c) views of the CUPID-0 detector. d) 3D view of the \CupidZ\ detector hosted in the \CuoreZ\ cryostat.}
\label{fig:setup}
\end{centering}
\end{figure}
The scintillation light emitted by each \EnrZnSe\ will be measured by two light detectors~\cite{Beeman:2013zva}, facing the top and bottom surfaces of the crystal (see Figure~\ref{fig:setup}-a). 
The light detectors are also bolometers made of disk-shaped Ge crystals (4.4\,cm diameter) grown by UMICORE using the Czochralski technique.
Since the energy emitted as scintillation light is only a few $\%$ of the total, the Ge crystals must be very thin to ensure a low heat capacity and, thus, a higher signal. For this reason, the light detectors were cut in 170\,$\mu$m thick wafers, that were polished and etched on both sides. The face showing the best optical properties was coated with a thin layer (70\,nm) of SiO in order to increase the light absorption~\cite{Beeman:2012cu,Mancuso:2014paa}.

To convert temperature variations into readable voltage signals, each ZnSe and Ge crystal will be equipped with a Neutron Transmutation Doped (NTD) Ge thermistor using a semi-automatic gluing system.
In addition, a Si Joule heater will be glued to the crystals for the offline correction of thermal drifts by heat pulses injection~\cite{Arnaboldi:2003yp}.

The \EnrZnSe\ and Ge bolometers will be mounted in an Oxygen Free High Conductivity (OFHC) copper structure, serving as thermal bath to cool the detectors at $\sim$10\,mK.
To account for different thermal contractions of the detectors and copper, the crystals will be secured to the copper frames using PTFE elements, acting also as weak thermal coupling to the bath.
The final layout of the \CupidZ\ detector, that will be hosted in the same cryostat used for \CuoreZ\ with an upgrade for the reduction of the microphonic noise, is shown in Figure~\ref{fig:setup}.

To test the \CupidZ\ assembly line, as well as to verify the compliance of the final detectors with the requirements on energy resolution, background rejection capability and intrinsic radio-purity,
we mounted an array of three \EnrZnSe\ crystals and performed a bolometric run.
In this paper we present the results of the test run and discuss the perspectives for \CupidZ.


\section{Detector}
\label{sec:detector}
The detector consists of three \EnrZnSe\ crystals (for a total \EnrZnSe\ mass of about 1.32\,kg) and four light detectors, arranged in a single tower following the layout depicted in Figure~\ref{fig:setup}.

Apart from the number of crystals, there are only few differences with the final \CupidZ\ protocol for the detector assembly: (1) the cleaning procedure defined for the surface of the crystals in \CupidZ\ was not yet applied the present crystals, as we were interested mainly in the bolometric performance and crystal bulk contaminations, (2) the detector was mounted above-ground, shortly after the crystal growth, while the \CupidZ\ towers will be assembled in a Radon-free underground clean room using crystals stored underground to reduce the cosmogenic activation, (3) during the measurement the \EnrZnSe\ were permanently exposed to smeared $\alpha$ sources of $^{147}$Sm (Q-value$\sim$2.3\,MeV) to assess the background rejection capability in a short measurement time.

\begin{figure}
\centering
  \includegraphics[width=0.25\textwidth]{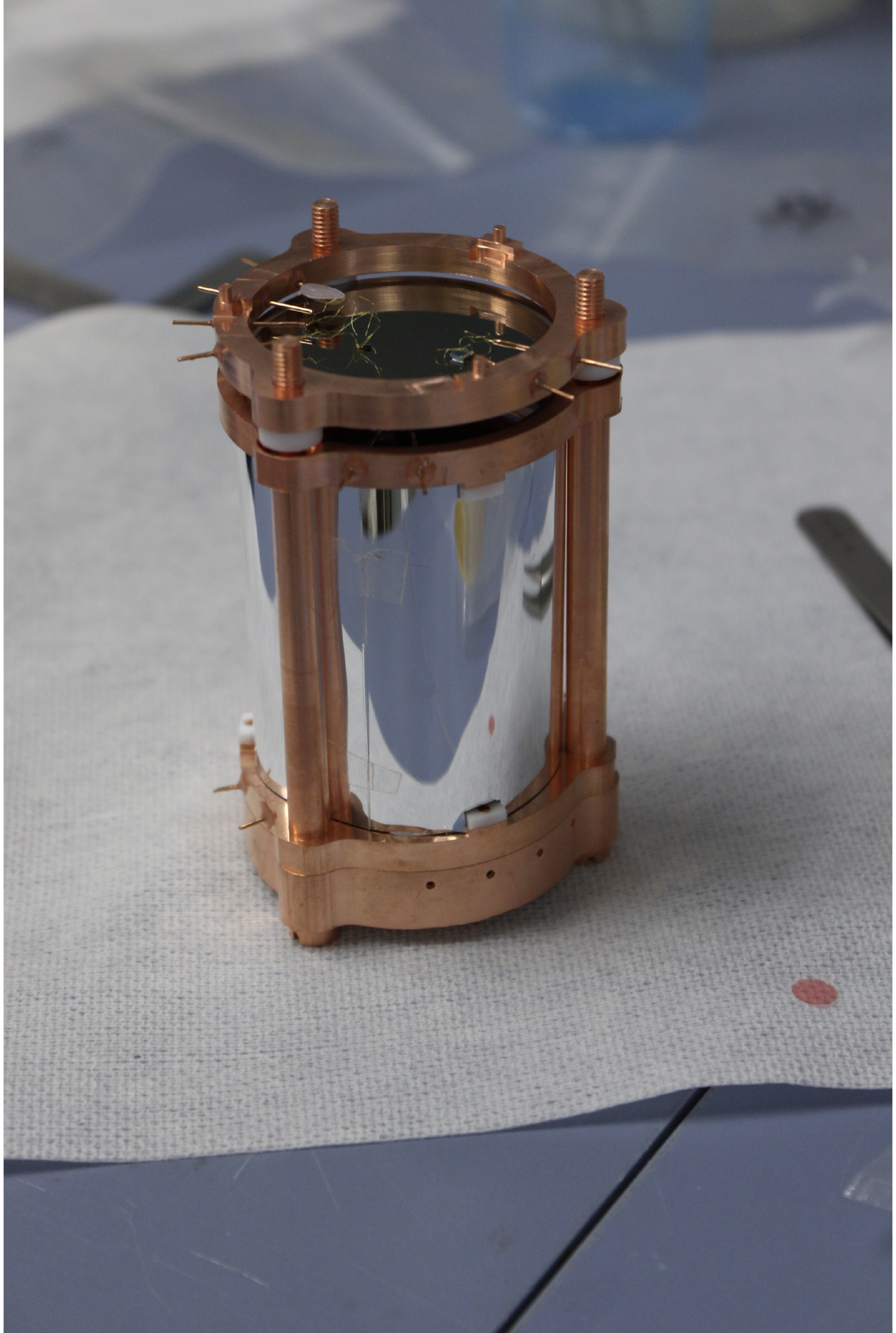}
   \includegraphics[width=0.2115\textwidth]{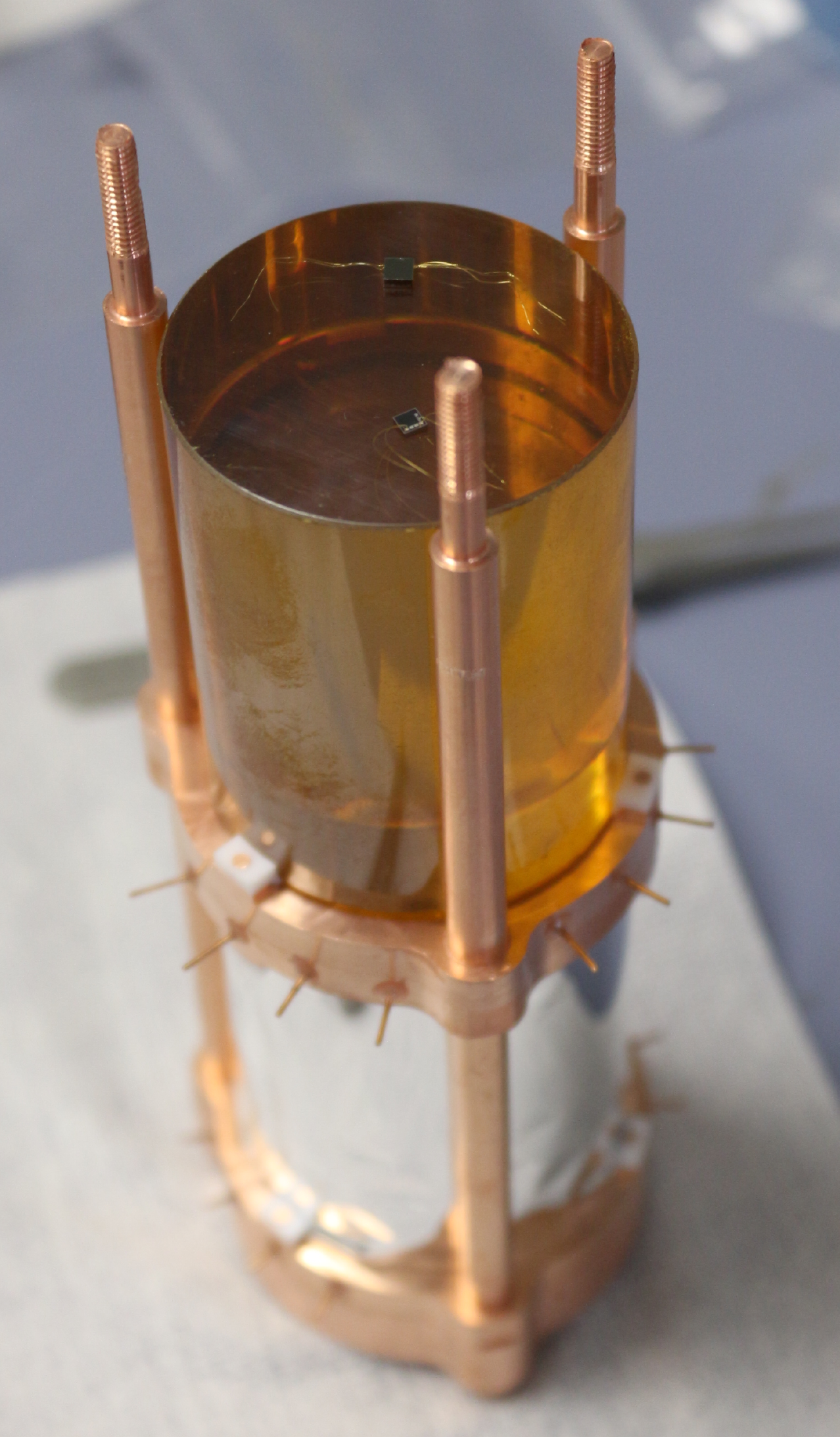}
\caption{Two stages of the tower assembly. Left: a light detector equipped with an NTD Ge thermistor and a Si Joule heater is mounted on top of a \EnrZnSe\ bolometer, surrounded by a 3M VM2002 reflecting foil. Right: the second \EnrZnSe\ bolometer is placed on top of the previous light detector.}
\label{fig:detector1}       
\end{figure}

We used two classes of NTD Ge thermistors, characterized by different dimensions: the large sensors (2.8$\times$3.0$\times$1.0\,mm) were attached on \EnrZnSe\ bolometers, while the smaller ones (2.8$\times$2.0$\times$0.5\,mm) were glued to the light detectors that, because of their much smaller mass, require sensors with lower heat capacity. 
Two stages of the detector assembly are shown in Figure~\ref{fig:detector1}. 

The \EnrZnSe\ array was anchored to the coldest point of a $^3$He/$^4$He dilution refrigerator located in the Hall C of LNGS. 
For technical reasons, the refrigerator could not be operated below 20\,mK, which is however satisfactory for our purposes. 
The final CUPID-0 towers will reach a much lower base temperature (about 10\,mK) exploiting the cryostat that was used for the \CuoreZ\ experiment. 

\begin{figure}
\centering
  \includegraphics[width=0.45\textwidth]{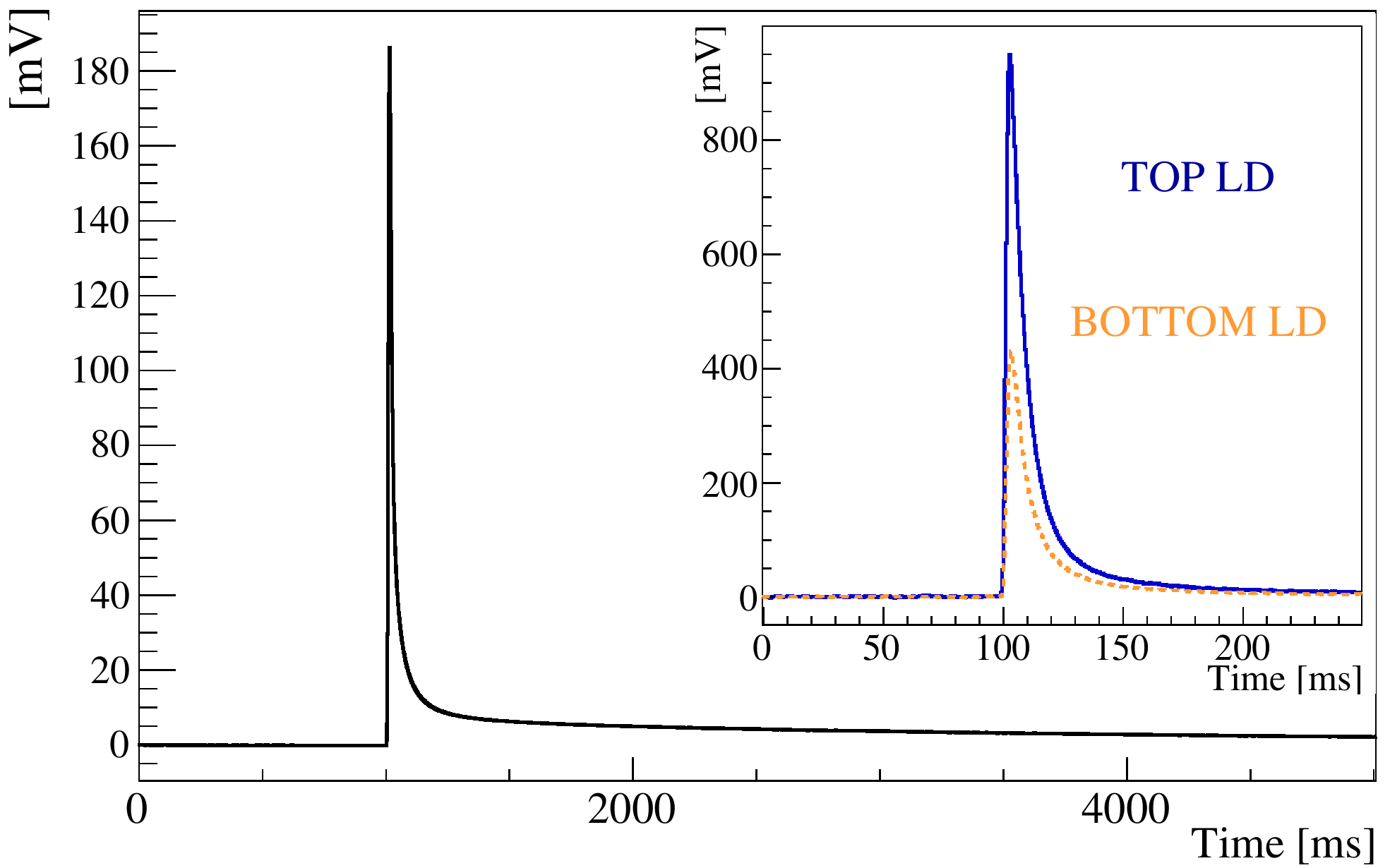}
\caption{Response of \EnrZnSe\ bolometers obtained by averaging tens of pulses with energy of 2615\,keV. Inset: corresponding scintillation light detected by the top (blue, continuous line) and bottom (orange, dotted line) light detectors. The side of the top light detector facing the \EnrZnSe\ bolometer was coated with SiO.}
\label{fig:detector2}    
\end{figure}

The signals were amplified by high gain voltage amplifiers with input stages based on Silicon JFETs, filtered by a 6-pole low-pass Bessel-Thomson filter~\cite{Arnaboldi:2006mx,Arnaboldi:2004jj,AProgFE} with roll-off rate of $120\un{dB/decade}$, and fed into an 18 bit NI-6284 PXI ADC unit. Due to the different time development of heat and light pulses (shown in Figure~\ref{fig:detector2}) we set a cut-off frequency of 63\,Hz for \EnrZnSe\ and 200\,Hz for light detectors.
When the software trigger of each \EnrZnSe\ fired, we saved on disk 5\,s long time-windows sampled at 1\,kHz. The light detectors were acquired with a shorter window of 250\,ms and a higher sampling frequency (2\,kHz). 
In addition to the derivative trigger, we used a second trigger that forced the acquisition of the light detectors every time the trigger of a \EnrZnSe\ fired.

To maximize the signal-to-noise ratio, the waveforms were processed offline with the optimum filter algorithm~\cite{Gatti:1986cw,Radeka:1966}.
Unlike the classical nuclear pulse shape processing (gaussian, semi-gaussian, CR-(RC)$^n$ etc.) the optimum filter estimates the signal at each frequency with weights equal the signal-to-noise ratio at the same frequency, 
embedding in its shape the proper timing parameters.
In our case it behaves as a pass-band filter with the lower frequency optimized for 1/f noise rejection, while the higher frequency is limited by signal roll-off, minimizing the contribution of white noise. 
Since light pulses feature a worse signal-to-noise ratio, we used a dedicated algorithm, exploiting the known time-delay between heat and light signals, to compute their amplitude~\cite{Piperno:2011fp}.

The general features of the \EnrZnSe\ bolometers are reported in Table~\ref{tab:generalfeatures}.
\begin{table}
\centering
\caption{Resistance of the NTD Ge thermistor in working conditions R$_{work}$, FWHM noise energy resolution after the optimum filter, voltage signal (in $\mu$V) produced by an energy release of 1\,MeV.}
\label{tab:generalfeatures}       
\begin{tabular}{lccc}
\hline\noalign{\smallskip}
                            &R$_{work}$      &baseline noise  &Response               \\
                            &M$\Omega$     &[keV FWHM]     &[$\mu$V/MeV]         \\
\hline
Zn$^{82}$Se-1     &0.20		     & 7.0                  &7.1                         \\
\hline
Zn$^{82}$Se-2     &0.22		      &14.1                &3.7                        \\
\hline
Zn$^{82}$Se-3     &0.17		     &18.6                 &2.9                          \\
\noalign{\smallskip}\hline
\end{tabular}
\end{table}
The working resistance of the NTD Ge thermistor R$_{work}$ depends on the detector temperature: the lower the temperature, the larger the resistance. 
The electronics channels are optimized for R$_{work}$ of the order of tens/hundreds of M$\Omega$ (corresponding to a crystal temperature lower than 10\,mK) but the high base temperature of the cryostat prevented the achievement of such a resistance, somehow spoiling the detector performance.
Given the low values of R$_{work}$ (reported in Table~\ref{tab:generalfeatures}), we estimated a noise contribution from the electronics of 7.0\,keV FWHM for Zn$^{82}$Se-1, 14.3\,keV for Zn$^{82}$Se-2 and 18.3\,keV for Zn$^{82}$Se-3. 
These values are in full agreement with the measured energy resolution of the detector baseline (Table~\ref{tab:generalfeatures}), leading to the consideration that a lower cryostat temperature would have provided a much better baseline resolution, as observed in other ZnSe prototypes~\cite{Beeman:2013vda}. 
The analysis of the noise power spectra of the detectors allowed to infer that the degradation of the baseline resolution could be mainly ascribed to the series noise (both 1/f and white) of the very front-end system, operated in unmatched conditions.
Nevertheless, our primary interest was the study of the detector performance at high energy where, as shown in the next section, the electronics noise is negligible. Furthermore, also the response of the bolometer, defined as the voltage signal produced by an energy deposit of 1 MeV, was affected by the higher than usual detector temperature. A lower temperature of operation, to be expected in the next runs, will allow to increase the energy conversion gain of the detectors, providing an enhancement of the baseline resolution. 
To face both the situations, low temperature with large detector impedance and warm temperature with small detector impedance, we developed a new set of preamplifiers able to match the latter condition with a noise smaller by a factor of 2.5 and input capacitance larger by a similar factor with respect to the actual setup.

\section{Energy Resolution}
\label{sec:resolution}
\begin{figure}
\centering
  \includegraphics[width=0.45\textwidth]{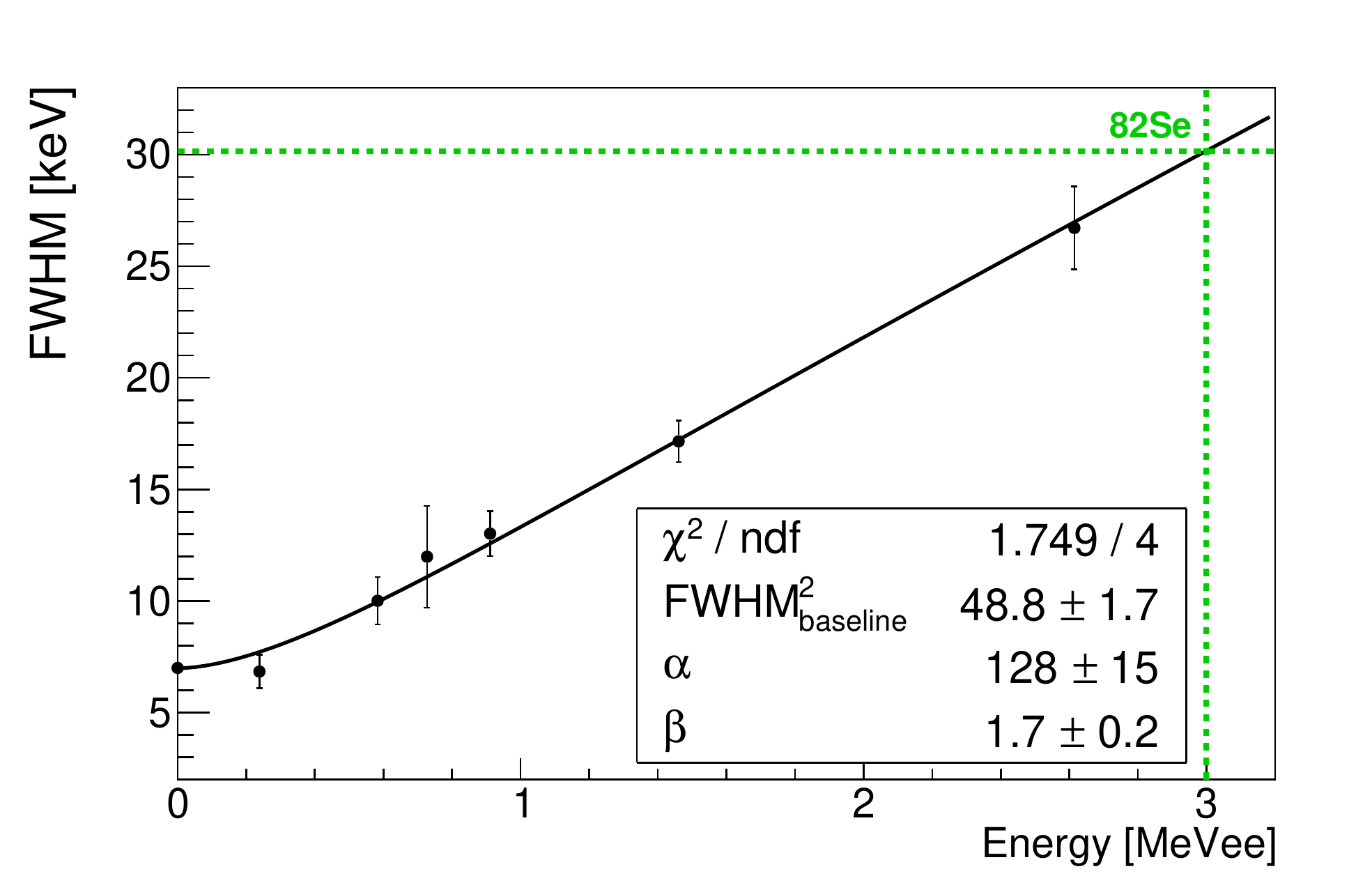}
\caption{FWHM energy resolution as a function of the energy (Zn$^{82}$Se-1) for the most intense $\gamma$ peaks produced by $^{228}$Th and $^{40}$K sources. The point at zero energy is the baseline energy resolution reported in Table~\ref{tab:generalfeatures}. The black line is the fit function: FWHM$^2$(E) = FWHM$^2_{baseline}$+$\alpha$E$^{\beta}$. The green dotted lines indicate the \Se\ Q-value.}
\label{fig:ZnSeresolution}    
\end{figure}

The \EnrZnSe\ bolometers were energy calibrated using $^{228}$Th and $^{40}$K sources, emitting $\gamma$ rays up to 2615\,keV.
We fitted the most intense $\gamma$ peaks to check the linearity of the detector response, as well as to determine the energy resolution (see Figure~\ref{fig:ZnSeresolution} for Zn$^{82}$Se-1).

Since the $\gamma$ natural radioactivity drops above the 2615\,keV line, we could not rely on calibration peaks at the Q-value of \Se. 
For this reason, we estimated the energy resolution in the ROI by fitting the energy resolution of the calibration peaks as a function of the energy with the following function:
\begin{equation}
FWHM^2(E) = FWHM^2_{baseline}+\alpha E^{\beta}
\end{equation}
where $FWHM^2_{baseline}$ accounts for the contribution of the electronic noise (see Table~\ref{tab:generalfeatures}), and $\alpha$ and $\beta$ are arbitrary coefficients.
At the \Se\ Q-value we derived a FWHM energy resolution of 30.1$\pm$1.7\,keV (Zn$^{82}$Se-1), 29.7$\pm$1.4\,keV (Zn$^{82}$Se-2) and 30.2$\pm$1.7\,keV (Zn$^{82}$Se-3), proving the reproducibility of the \EnrZnSe\ bolometers.

These values are about a factor 3 larger than the initial target of the experiment (10\,keV FWHM~\cite{Beeman:2013sba}). Even if a further cooling of the crystals is expected to improve the resolution, we will demonstrate in the next sections that \EnrZnSe\ bolometers with  30\,keV FWHM would anyway comply with the requirements of \CupidZ.

\section{Performance of Light Detectors}
The sensitivity of light detectors is crucial for \CupidZ, as the background rejection capability relies on the different light emission of electrons and $\alpha$ particles.
In this run, all the light detectors were permanently exposed to $^{55}$Fe X-rays sources (characterized by peaks at 5.9 and 6.4\,keV) for energy-calibration.
To further investigate the reproducibility of light detectors, we assembled a second array, with three Zn$^{nat}$Se interleaved by four Ge disks. 
Two of these light detectors were equipped with large NTD Ge thermistors, similar to those attached on Zn$^{nat}$Se crystals, to investigate the effect of the sensor heat capacity on the bolometric performance. 
The features of the light detectors are reported in Table~\ref{tab:generalfeaturesLD}.

As explained in section~\ref{sec:detector}, the high cryostat temperature could produce an electronic noise limiting the detector energy resolution. Nevertheless, it is worth highlighting that, in absence of any signal amplification, energy resolutions as good as  those reported in Table~\ref{tab:generalfeaturesLD} have never been achieved with wide area Ge light detectors equipped with conventional NTD Ge thermistors.
The detector response in terms of signal height, as well as the energy resolution, showed variations lower than 40$\%$ across the 6 light detectors equipped with small NTD Ge sensors, proving the reproducibility of the light detectors performance in view of \CupidZ.
These results can mainly be ascribed to the improvement of the interface between the Ge disks and the NTD Ge thermistors provided by dedicated surface treatments (etching and polishing), as well as by the semi-automatic gluing system that enhanced the reproducibility of the detectors features

\begin{table}
\centering
\caption{Features of light detectors equipped with small NTD Ge sensors: $\tau_{r}$ and $\tau_{d}$ are the rise and decay times, computed as the time difference between the 90$\%$ and 10$\%$ of the leading edge and as the time difference between the 30$\%$ and 90$\%$ of the trailing edge respectively. The other parameters were defined in Table~\ref{tab:generalfeatures}. In the last two lines we report the features of light detectors equipped with large NTD Ge sensors.}
\label{tab:generalfeaturesLD}       
\begin{tabular}{ccccc}
\hline\noalign{\smallskip}
R$_{work}$     &Response                   &baseline noise     &$\tau_{r}$ &$\tau_{d}$\\
\MO		       &[$\mu$V/MeV]            &[eV FWHM]           &[ms]         &[ms]\\
\hline
0.63		       &1.0$\times$10$^3$     &134			&1.7		&3.4 \\
1.45		       &1.6$\times$10$^3$     &92			&1.9		&5.7 \\
0.71		       &1.0$\times$10$^3$     &103			&1.8		&5.4 \\
0.89		       &1.7$\times$10$^3$     &76			&1.7		&5.1 \\
0.85		       &1.7$\times$10$^3$     &89			&1.7		&5.2 \\
0.72		       &1.1$\times$10$^3$     &108                       &1.8		&5.4 \\
\hline
0.65		       &1.2$\times$10$^3$     &260                       &1.8		&9.2 \\
0.37		       &0.6$\times$10$^3$     &160                       &1.9		&11.0 \\
\noalign{\smallskip}\hline
\end{tabular}
\end{table}


\section{Background Study}
\subsection{Cosmogenic Activation}
To investigate the cosmogenic activation of \EnrZnSe\ crystals we exploited the comparison between data extracted from the cosmogenic activation calculation software Activia~\cite{activia} and the measurements performed with this array. 
The isotope of (enriched) Selenium with the highest activation rate is $^{75}$Se (Q-value$\sim$864\,keV, T$_{1/2}\sim$118\,d) which decays via electron capture with a rather complex combination of de-excitation $\gamma$'s and X-rays.
None of the enriched crystals showed evidences of $^{75}$Se, and we set a 90$\%$ C.L. upper limit on its activity of 14\,$\mu$Bq/kg.
All the other isotopes are expected to have a much lower activation rate. Among them, the only emitters that could produce background events for the \DBD\ of \Se\ are $^{48}$V (Q-value $\sim$4012\,keV, T$_{1/2}\sim$16\,d) and $^{56}$Co (Q-value $\sim$4566\,keV, T$_{1/2}\sim$79\,d). Both these isotopes have activation rates about 3 orders of magnitude lower than $^{75}$Se. Furthermore, due to their short half-lives, they will not partecipate to the background of \CupidZ, as the crystals will be permanently stored underground until operations.

We performed a similar analysis to investigate the activation of Zn.
In this case, the isotope with highest activation rate is $^{65}$Zn (Q-value$\sim$1352\,keV and T$_{1/2}\sim$244.26\,d). The signature of $^{65}$Zn was observed in all the \EnrZnSe\ crystals with similar activities of 3.06$\pm$0.44\,mBq/kg. 
Also in this case, $^{48}$V and $^{56}$Co would be the only dangerous emitters produced by cosmogenic activations, with activation rates much lower than $^{65}$Zn.
As said before, due to the low activation and the short half-lives, these isotopes will not be of concern for \CupidZ. 

Nevertheless, since this test run was performed shortly after the crystal production and delivery, we expect $^{56}$Co to contribute about 1~$\gamma$/day in the total energy spectrum of the array above 2615\,keV.

\subsection{Alpha Background rejection}
\label{sec:alphabackground}
The large light output of \EnrZnSe\ crystals and the excellent performance of the cryogenic light detectors provided a very efficient $\alpha$ background rejection.
The three crystals featured a similar light yield (LY), defined as the amount of light (in keV) measured when an interaction of 1\,MeV occurs in the \EnrZnSe\ bolometer.
ZnSe crystals operated at cryogenic temperatures show a rather peculiar (and not yet fully understood) feature: the LY of electrons is lower with respect to the one of $\alpha$ particles.
With this array we measured LY$_{\beta/\gamma}$ ranging from 3.3 to 5.2\,keV/MeV, to be compared with LY$_{\alpha}$ ranging from 9.1 to 14.1 keV/MeV.

\begin{figure}
\centering
  \includegraphics[width=0.45\textwidth]{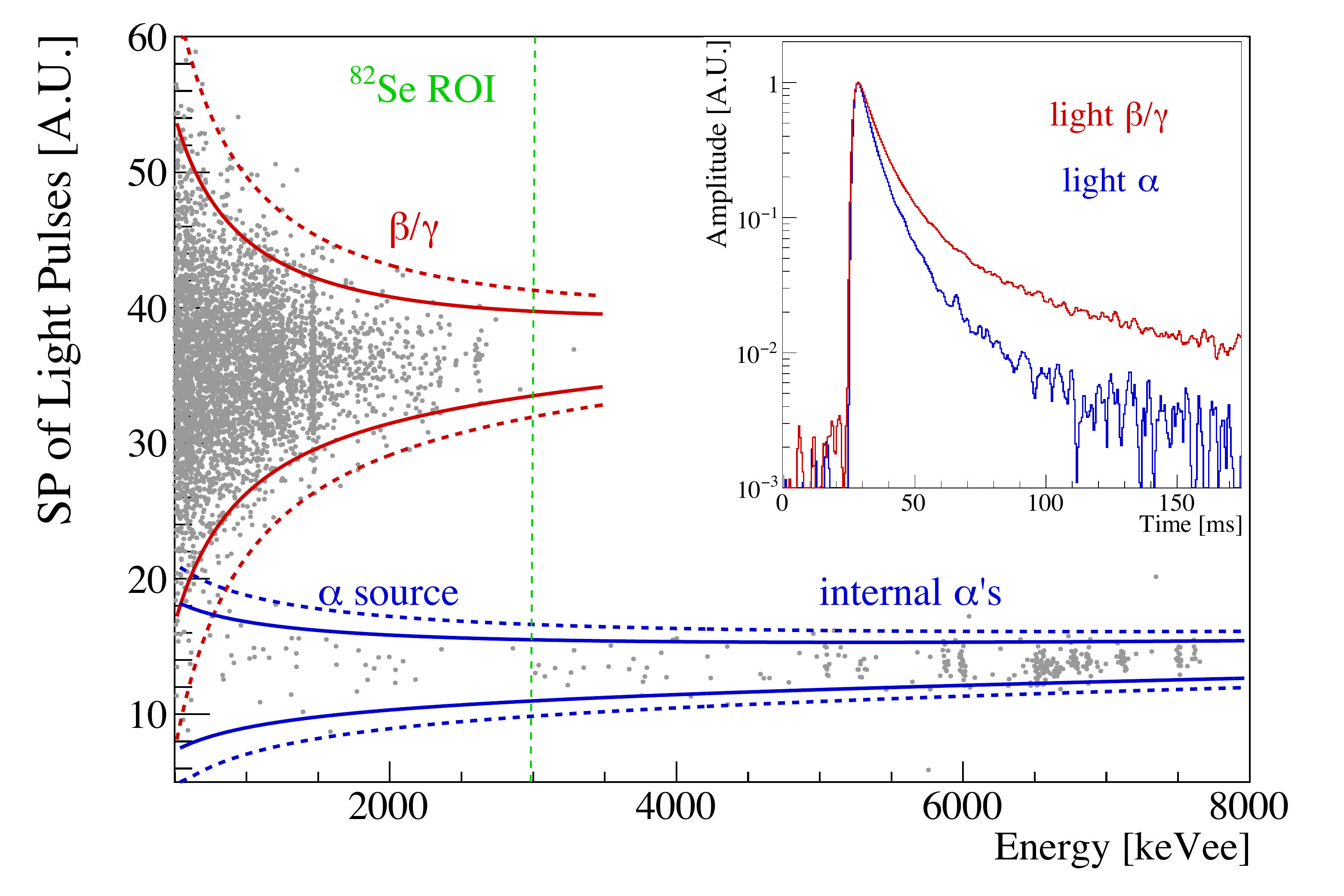}
\caption{Shape parameter of a light detector as a function of the energy released in Zn$^{82}$Se-1 ($\sim$530\,h). The red and blue lines indicate the 2$\sigma$ (continous) and 3$\sigma$ (dotted) $\beta/\gamma$ and $\alpha$ bands respectively. $\alpha$ events produced by the smeared Sm source (below 3\,MeV$_{ee}$) and by contaminations of the crystal bulk (peaks above 5\,MeV$_{ee}$) can be easily rejected, in particular in the region of interest for the \Se\ \DBD\ (green lines). The other \EnrZnSe\ showed similar results. Inset: time development of light pulses produced by  $\beta/\gamma$ (blue) and $\alpha$ (red) interactions with energy of about 2.6\,MeV. }
\label{fig:TVR}    
\end{figure}

Even if the different LY allows to identify and separate electrons and $\alpha$ events, we used a more powerful estimator for background rejection: the shape of the light pulses.
As shown in the inset of Figure~\ref{fig:TVR}, the light pulses produced by $\beta/\gamma$ interactions are slower than light pulses produced by $\alpha$'s of the same energy.

To exploit this difference, we used a shape-sensitive parameter (SP) computed on the optimum filtered pulse and defined as:
$$
SP = \frac{1}{Aw_r}\sqrt{\sum_{i=i_M}^{i_M+\omega_r}(y_i-As_i)^2}
$$
where y$_i$ is the pulse, A and i$_M$ its amplitude and maximum position, s$_i$ the ideal signal pulse scaled to unitary amplitude and aligned to y$_i$, $w_r$ the right width at half maximum of s$_i$. 
The Shape Parameter of light pulses computed for the best light detector is shown for Zn$^{82}$Se-1 in Figure~\ref{fig:TVR}.

Because of the worse signal-to-noise ratio, the $\beta/\gamma$ and $\alpha$ bands become wider at lower energies.
To compute the discrimination capability at the \DBD\ energy, we divided the $\alpha$ and $\beta/\gamma$ bands in intervals and made Gaussian fits to derive the mean value ($\mu$) and the standard deviation  ($\sigma$) of the Shape Parameter in each interval. We fitted the energy dependence of  $\mu$(E) and $\sigma$(E) with polynomial functions and defined the Discrimination Power (DP) as a function of the energy as:
$$
DP(E) = \frac{\left|\mu_{\alpha}(E)-\mu_{\beta\gamma}(E)\right|}{\sqrt{\sigma_\alpha^2(E)+\sigma_{\beta\gamma}^2(E)}}
$$
and found $DP=12$ at the \Se\ Q-value. The same analysis was made on the other two crystals, obtaining a DP of 11 and 10 for Zn$^{82}$Se-2 and Zn$^{82}$Se-3 respectively.

\begin{figure}
\centering
  \includegraphics[width=0.5\textwidth]{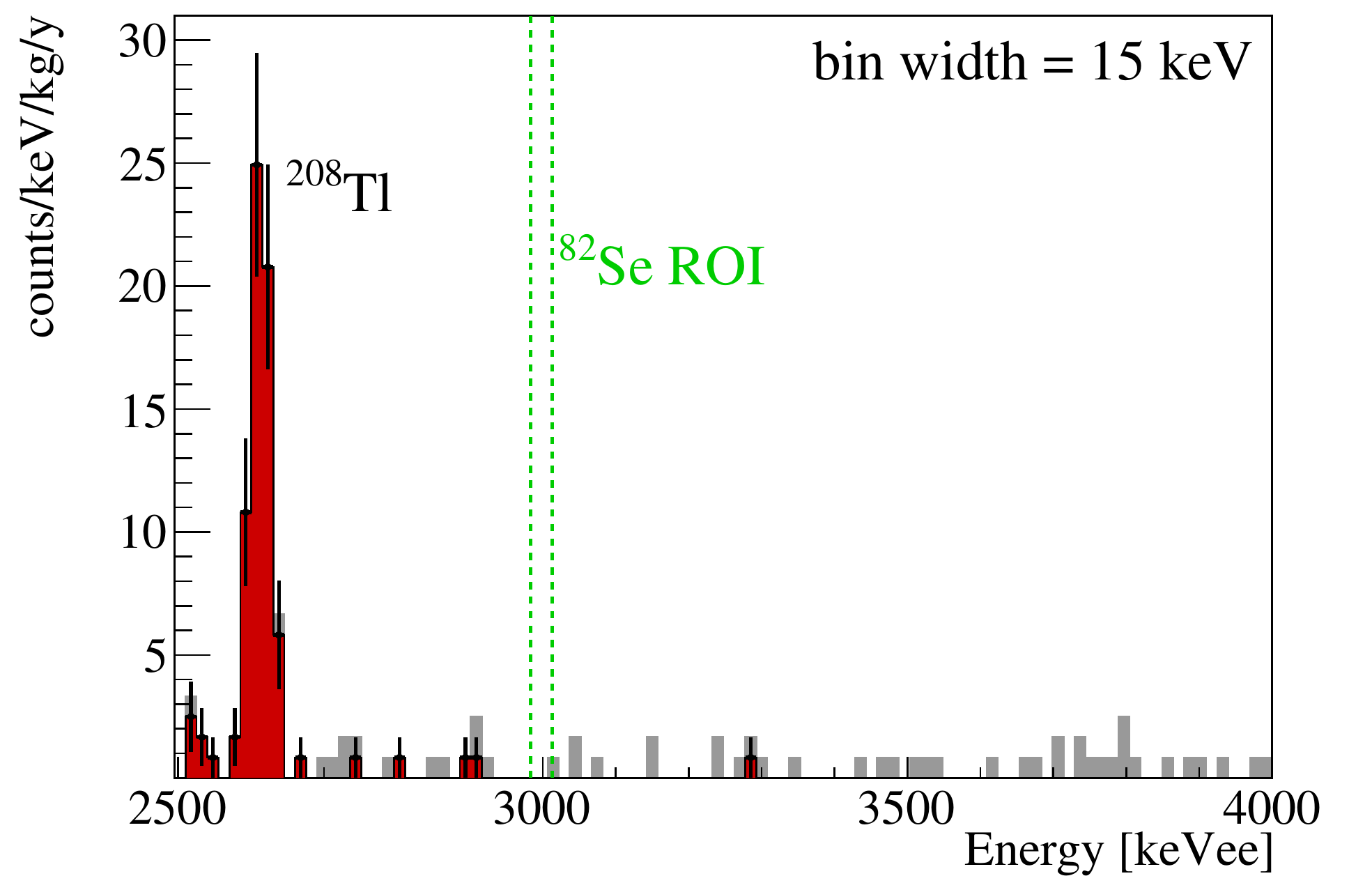}
\caption{High-energy spectrum of the \EnrZnSe\ array collected in 530\,h before (gray) and after (red) the rejection of the  $\alpha$ background. The green lines indicate a FWHM region centered on the \Se\ Q-value.}
\label{fig:backgroundROI}    
\end{figure}

We report in Figure~\ref{fig:backgroundROI} the zoom at high energy of the spectrum collected with the \EnrZnSe\ array in about 530\,h, before and after the rejection of the $\alpha$ background and of the events occurring simultaneously in more than one crystal.
The number of events above the $^{208}$Tl peak is very small if we consider that the cryogenic set-up used for this test did not provide an efficient shield against $^{214}$Bi produced in Rn-contaminated air, and that the crystals were measured immediately after production (when the activation could still give rise to an important background).
For \CupidZ\ we foresee an even better result, as the cryostat used by \CuoreZ\ will be equipped with a more effective Rn-free system and a more efficient shielding.
Moreover, storing the crystals underground will suppress the activity of isotopes like $^{56}$Co (T$_{1/2}\sim$79\,d) that, as explained in the previous section, was expected to produce a non-negligible background in this run.


\subsection{Crystal Contaminations}
The radio-purity of \EnrZnSe\ crystals is of primary concern for the achievement of the target sensitivity, as the long-living isotopes of the natural $^{238}$U and $^{232}$Th chains may give rise to an irreducible background in the region of interest around 3 MeV.

Due to their very short range, $\alpha$ particles can give a clear indication of bulk or surface contaminations. If the $\alpha$-decaying isotope is located in the crystal bulk, both the $\alpha$ particle and the nuclear recoil are absorbed by the bolometer, giving rise to a  peak at the Q-value of the decay. On the contrary, if the contaminant is located on the crystal surface, $\alpha$'s (and with much lower probability nuclear recoils) may leave the crystal before being stopped and the event is recorded with an energy lower than the Q-value. Only for very shallow surface contamination the $\alpha$ line (about 100\,keV below the Q-value) becomes visible.

\begin{figure}
\centering
  \includegraphics[width=0.5\textwidth]{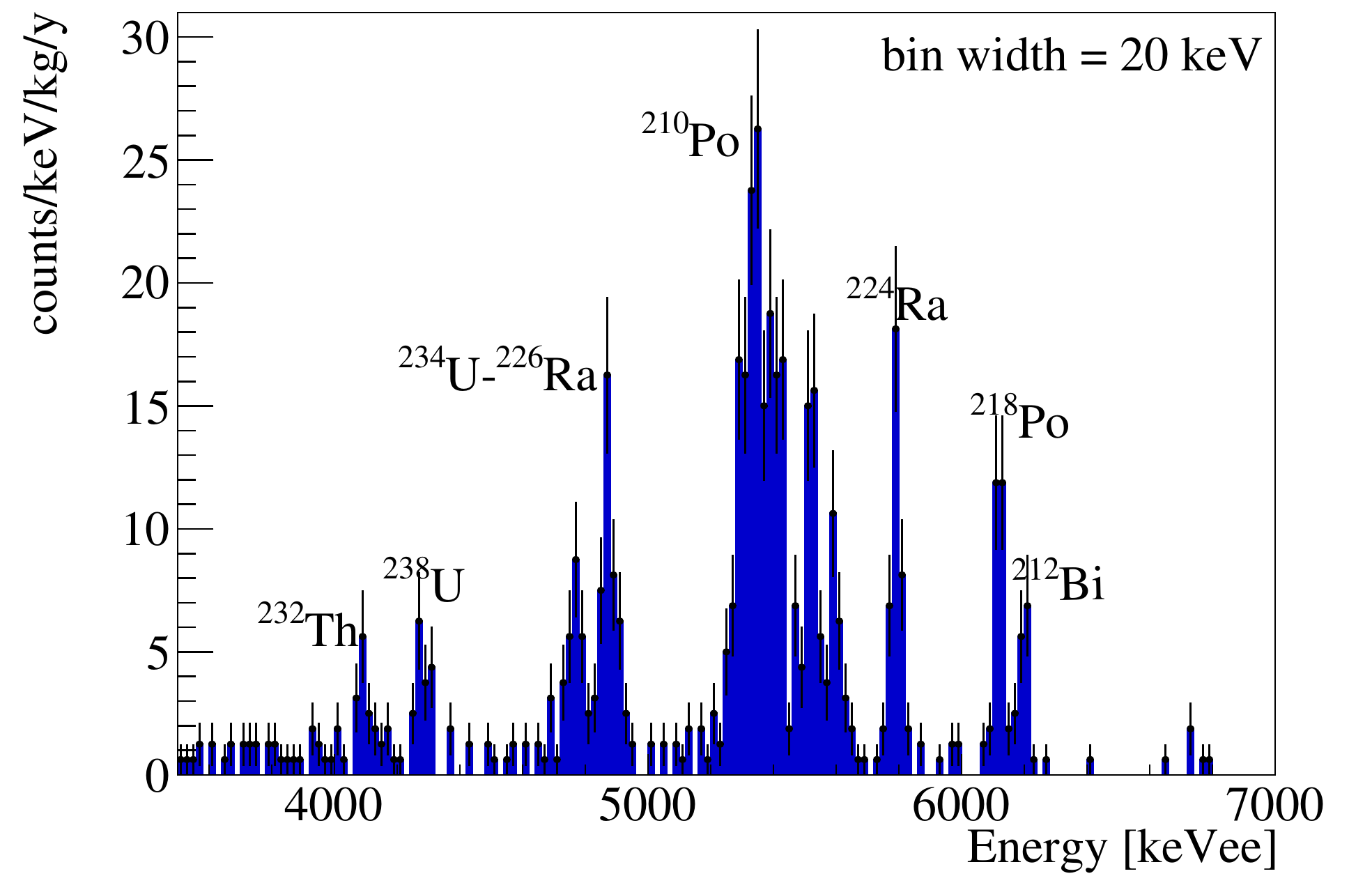}
\caption{$\alpha$ region of the \EnrZnSe\ array collected in 530\,h. The spectrum was energy-calibrated using the nominal energy of the most prominent $\alpha$ peaks.}
\label{fig:backgroundAlpha}    
\end{figure}

To extract the crystal contaminations in the isotopes of the $^{238}$U and $^{232}$Th chains, we analyzed the $\alpha$ region of the total energy spectrum, reported in Figure~\ref{fig:backgroundAlpha}. 
The same analysis was performed on the energy spectra of each Zn$^{82}$Se to highlight possible differences, and the results are shown in Table~\ref{tab:internalcontaminations}.
In some cases we had no evidences of contaminations, thus we reported the 90$\%$ C.L. upper limit.
For this purpose, we defined the signal as the number of events falling in the energy region [Q-value - 3$\sigma$, Q-value + 3$\sigma$] and the background as the average number of events falling in the 3$\sigma$ side-bands of this interval. Following the Feldman-Cousins approach, we computed the 90$\%$ C.L. upper limit on the number of events and, correcting for the branching ratio of the nuclide, we inferred the upper limit on the activity.
The $\alpha$ lines not quoted in the table are the two $\sim$6\,MeV lines of $^{232}$Th because they produce a pile-up event rejected as a deformed signal by pulse shape cuts, and the two $\alpha$ lines that in both chains are summed to the following beta emission (Bi-Po events) producing a continuum above 8 MeV.
\begin{table}
\centering
\caption{Crystal contaminations in the isotopes of $^{238}$U and $^{232}$Th chains derived from the analysis of the $\alpha$ region (see Figure~\ref{fig:backgroundAlpha}). We made the same analysis on the energy spectrum of each Zn$^{82}$Se bolometer to highlight differences among crystals. Isotopes with long half-lives, that can break the secular equilibrium, are highlighted in bold type.}
\label{tab:internalcontaminations}       
\begin{tabular}{lcccc}
\hline\noalign{\smallskip}
                              &Zn$^{82}$Se-1      &Zn$^{82}$Se-2        &Zn$^{82}$Se-3    &Array 	  \\
                              &[$\mu$Bq/kg]         &[$\mu$Bq/kg]          &[$\mu$Bq/kg]       &[$\mu$Bq/kg]\\
\hline
{\bf $^{232}$Th}     &13 $\pm$ 4            &13 $\pm$ 4              &$<$5                    &7 $\pm$ 2	  \\
\hline
{\bf $^{228}$Th}     &32 $\pm$ 7            &30 $\pm$ 6              &22 $\pm$ 4          &26 $\pm$ 2 \\
\hline
$^{224}$Ra            &29 $\pm$ 6            &26 $\pm$ 5              &23 $\pm$ 5          &27 $\pm$ 3 \\
\hline
$^{212}$Bi              &31 $\pm$ 6            &31 $\pm$ 6             &23 $\pm$ 5          &29 $\pm$ 3 \\
\hline
\hline
{\bf $^{238}$U}       &17 $\pm$ 4            &20 $\pm$ 5              &$<$10                 &10 $\pm$ 2   \\
\hline
{\bf $^{234}$U+$^{226}$Ra}       &42 $\pm$ 7             &30 $\pm$ 6              &23 $\pm$ 5                &33 $\pm$ 4   \\
\hline
{\bf $^{230}$Th}     &18 $\pm$ 5             &19 $\pm$ 5             &17 $\pm$ 4         &18 $\pm$ 3  \\
\hline
$^{218}$Po            &20 $\pm$ 5             &24 $\pm$ 5              &21 $\pm$ 5        &21 $\pm$ 2   \\
\hline
{\bf $^{210}$Pb}     &100 $\pm$ 11         &250 $\pm$ 17           &100 $\pm$ 12    &150 $\pm$ 8   \\
\noalign{\smallskip}\hline
\end{tabular}
\end{table}

The energy spectrum reported in Figure~\ref{fig:backgroundAlpha} was reconstructed using a Monte Carlo simulation (not shown),
obtaining activities compatible with those reported in Table~\ref{tab:internalcontaminations}.

The position and shape of the $\alpha$ peaks suggested that most of the contaminants are uniformly diffused in the crystal bulk, 
but we could not completely exclude the hypothesis that part of the contaminations is located deep in the crystal surface. 
For simplicity, however, the activities reported in Table~\ref{tab:internalcontaminations} were normalized to the mass of the crystals.

The only isotope showing a clear evidence for a surface contamination was $^{210}$Po that, because of its rather short half-life ($\sim$138\,d) was likely produced by $^{210}$Pb.
As shown in Figure~\ref{fig:backgroundAlpha}, $^{210}$Pb produced two structures at $\sim$5.3\,MeV ($\alpha$) and $\sim$5.4 MeV ($\alpha$ + nuclear recoil).
The presence of the first peak could be explained by the simulation only assuming a shallow contamination of the crystal surface or of the inert materials surrounding the detector.
The activity of $^{210}$Pb shown in Table~\ref{tab:internalcontaminations} was computed attributing the entire event rate (surface + bulk) to a contamination of the crystal bulk.
The presence of this contaminant, however, is not of concern for \CupidZ, as none of the the daughters of $^{210}$Pb/$^{210}$Po  produces dangerous $\beta/\gamma$ events.

In contrast to $^{210}$Po, the signature of isotopes belonging to $^{238}$U and $^{232}$Th chains could be explained both by crystal bulk contaminations, and by deep contaminations of the crystal surface ($>$\,0.1$\mu$m).
If the contaminations were located in the crystal surface, we would expect a background reduction after the surface processing. The \EnrZnSe\ crystals, indeed, will be treated with a cleaning procedure similar to the one developed for TeO$_2$ crystals by the CUORE collaboration~\cite{ioanprod}, that proved to be very effective.

Finally, the GEANT-4 simulation allowed to infer the expected background contribution to the region of interest for \CupidZ.
We implemented the geometry described in Figure~\ref{fig:setup}, and assumed a discrimination power of 12 (see section~\ref{sec:alphabackground}), and an energy resolution of 30\,keV at the Q-value of \Se\ (see section~\ref{sec:resolution}).

If all the contaminants were diffused in the crystal bulk, we would expect a background at the \Se\ Q-value of 4$\times$10$^{-2}$\,counts/keV/kg/y.
Since \DBD\ electrons are expected to be contained in a single crystal (with a containment efficiency of 80$\%$) while the events produced by $^{208}$Tl decay mainly occur in more than one detector,
we can exploit coincidences  among nearby crystals to suppress this background to the level of 2.3$\times$10$^{-2}$\,counts/keV/kg/y.
Taking advantage of the small time delay between $^{208}$Tl and its mother ($^{212}$Bi), we can reach a level of 1$\times$10$^{-3}$\,counts/keV/kg/y with a negligible dead-time (2$\%$)~\cite{Beeman:2013vda,Beeman:2011bg},
in compliance with the requirements of \CupidZ.

If all the contaminants were located only on the crystal surface, and in the unlikely event that crystal polishing does not change crystal surface activity, the background in the region of interest would be of 6$\times$10$^{-3}$\,counts/keV/kg/y (coincidence suppression included), dominated by $^{208}$Tl interactions. 
In this case, the background suppression obtained exploiting the time delay between $^{208}$Tl and $^{212}$Bi would be slightly less effective, and would allow to reach the level of 3$\times$10$^{-3}$\,counts/keV/kg/y. 
However, following what observed in CUORE~\cite{Alessandria:2011vj,Pavan:2015lmm}, crystal surface treatment allows a reduction of surface activity by more than a factor 6. In this case the expected counting rate could be even lower than 10$^{-3}$\,counts/keV/kg/y.


\section{Conclusions and Perspectives}
The results presented in this paper allowed to assess the performance of the first \EnrZnSe\ array in view of \CupidZ.
We demonstrated that the assembly line guarantees the reproducibility both of light detectors, characterized by an unprecedented sensitivity, and of  \EnrZnSe\ bolometers.
We derived the energy resolution of \EnrZnSe\ detectors, obtaining 30\,keV FWHM at the Q-value of \Se\ \DBD, and we proved that the excellent performance of light detectors allows to completely disentangle and reject the background due to $\alpha$ interactions.

\CupidZ\ is expected to run at least for 1\,y of live-time to prove the potential of the scintillating bolometers technology, and the high number of emitters will allow to reach a remarkable sensitivity on the \Se\ \DBD.
Thanks to the low crystal contaminations, the expected background should be lower than 1.5$\times$10$^{-3}$\,counts/keV/kg/y, mainly produced by the cryogenic setup~\cite{Beeman:2013sba}.
As explained in Ref~\cite{Cremonesi:2012av}, when the background is low enough to produce a number of events in the ROI of the order of unity along the experiment life (``zero background'' approximation), the sensitivity scales as:
$$
S_{1/2}^{0\nu} = -\frac{ln(2)}{ln(1-C.L./100)}\frac{N_{A} a \eta}{W}\epsilon \cdot M\cdot t \cdot f(\delta E)
$$
where C.L. is the confidence level, 
$N_{A}$ the Avogadro constant, 
$a$ the isotopic abundance, 
$\eta$ the stoichiometric coefficient of the \DBD\ candidate, 
$W$ the atomic weight. 
$\epsilon$ is the efficiency (80$\%$ for these \EnrZnSe), 
M the detector active mass, 
t the live-time 
and $f(\delta E)$ the fraction of signal event that fall into the considered energy region (0.76 when $\delta$E is the detector FWHM energy resolution).

Given a background index of  1.5$\times$10$^{-3}$\,counts/keV/kg/y, we expect $\sim$0.6 events in 1 year in a FWHM energy interval centered around the \Se\ Q-value.
With a lower cryostat temperature, we expect \CupidZ\ to feature a better energy resolution and, as a consequence, an even lower number of background events in the region of interest.
Thus, the sensitivity can be computed in the ``zero background'' approximation and results 9.3$\times$10$^{24}$\,y at 90$\%$ C.L. in 1 year of data taking. The sensitivity will increase linearly with the live-time of the experiment as long as the conditions will be compatible with the zero background approach.

\begin{acknowledgements}
This work was partially supported by the LUCIFER experiment, funded by ERC under the European Union's Seventh Framework Programme (FP7/2007-2013)/ERC grant agreement n. 247115, funded within the ASPERA 2nd Common Call for R\&D Activities. We are particularly grateful to M.~Iannone for its help in all the stages of the detector construction, to M.~Guetti for the assistance in the cryogenic operations and to the mechanical workshop of LNGS (in particular E.~Tatananni, A.~Rotilio, A.~Corsi, and B.~Romualdi) for continuous and constructive help in the overall set-up design.
\end{acknowledgements}

\bibliographystyle{spphys}       

\end{document}